\documentclass{aastex62}

\submitjournal{AAS}

\shorttitle{TDVs}
\shortauthors{Boley et al.}

\begin{document}

\title{Transit Duration Variations in Multi-Planet Systems}
\correspondingauthor{Aaron C. Boley}
\email{aaron@aaronboley.com}

\author{Aaron C.~Boley}
\affil{The University of British Columbia \\
6224 Agricultural Road \\
Vancouver, BC V6T 1Z1, Canada}

\author{Christa Van Laerhoven}
\affil{The University of British Columbia \\
6224 Agricultural Road \\
Vancouver, BC V6T 1Z1, Canada}

\author{A.~P. {Granados Contreras}}
\affil{ASIAA \\
AS/NTU Astronomy-Mathematics Building, No.1, Sec. 4\\
Roosevelt Rd, Taipei 10617, Taiwan, R.O.C.}

\begin{abstract}

A planet's orbital orientation relative to an observer's line of sight determines the chord length for a transiting planet, i.e., the projected distance a transiting planet travels across the stellar disc. For a given circular orbit, the chord length determines the transit duration. 
Changes in the orbital inclination, the direction of the longitude of ascending node, or both, can alter this chord length and thus result in transit duration variations (TDVs).  
Variation of the full orbital inclination vector can even lead to de-transiting or newly transiting planets for a system.  
We use Laplace-Lagrange secular theory to estimate the fastest nodal eigenfrequencies for over 100 short-period planetary systems.  The highest eigenfrequency is an indicator of which systems should show the strongest TDVs.
We further explore five cases (TRAPPIST-1, Kepler-11, K2-138, Kepler-445, and Kepler-334) using direct N-body simulations to characterize possible TDVs and to explore whether de-transiting planets could be possible for these systems.  
A range of initial conditions are explored, with each realization being consistent with the observed transits. 
We find that tens of percent of multiplanet systems have fast enough eigenfrequencies to expect large TDVs on decade timescales.  
Among the directly integrated cases, we find that de-transiting planets could occur on decade timescales and TDVs of 10 minutes per decade should be common.

\end{abstract}

\keywords{Exoplanets, Exoplanet evolution, Exoplanet dynamics, Transit duration variation method}

\section{Introduction} \label{sec:intro}

{\it Kepler} has revealed that short-period multiplanet systems are ubiquitous in the Galaxy \citep{burke_etal_2014}.  
Recent work suggests that 30\% of solar-type stars have Kepler-like systems with planets on orbits less than 400 d, with an average multiplicity of three planets per system \citep{zhu_etal_2018}.  
M dwarfs are predicted to have even higher occurrence rates \citep{mulders_etal_2015}.
However, we do not know whether the many known Systems with Tightly-packed Inner Planets (STIPs, i.e., the short-period multiplanet systems) harbour additional planets at larger stellarcentric distances, loosely resembling the Solar System's configuration, nor do we know in general the full 3D orientation of the observed planets' orbits.  

STIPs have already shown that planetary types and system configurations are diverse,  providing new information for evaluating formation and evolution paradigms, such as disc migration \citep{kley_nelson_2012}, N-body dynamical migration, at least for hot Jupiters \citep{fabrycky_tremaine_2007,wu_lithwick_2011,petrovich_2015}, and {\it in situ} formation/assembly \citep{hansen_murray_2012,chatterjee_tan_2014,boley_etal_2014}. Knowing whether STIPs also contain long period planets (or not) is of keen interest for interpreting the existing data \citep[see, e.g., ][]{zhu_wu_2018} and providing context to planet formation theory.
As an example, some broad classes of formation models suggest that STIPs are unlikely to have extensive outer planetary systems \citep{ormel_etal_2017}.
The full 3D planetary orbits are needed to better understand the actual inclination distribution of planets (with formation implications), and the evolution of those orbits, in turn, can constrain the perturbations acting on the planets. It is in this spirit that we focus on potential observable changes to transits in a given STIP due to the secular evolution of planetary orbits.

Planetary transit durations depend on several factors, including the planet's orbital period, the size of the star and planet, and the projection of the planet's orbit as seen by the observer.  
The projected path of the transiting planet across the stellar disc is the transit chord.
If the effective size of the star is known, then the duration of a transit can be used to infer the chord length and the impact parameter $b$, i.e., the offset of the chord from the centre of the stellar disc.  

The impact parameter is related to the inclination $i$ of the planet's orbit {\it relative to the observer} by $b= \frac{r}{R} \cos i$, where $r$ is the star-planet separation during the transit and $R$ is the effective size of the star. 
By convention, $i=90^\circ$ means the orbit normal lies in the skyplane, while for $i=0^\circ$,  the orbit normal is parallel to the observer's line of sight.  

Although we can infer the orientation of a planetary orbit relative to the observer, the transit duration does not in general tell us the {\it orbital} inclination $I$ of a planet relative to any given reference plane. 
Ultimately, the orbital inclination is a vector $\vec{I}$, with a direction toward the longitude of ascending node $\Omega$ (or simply the node) and magnitude $I$.  
As such, one cannot relate $i$ to $I$ in a general way, complicating the interpretation of multiplicity statistics and typical planetary inclinations \citep{tremaine_dong_2012}.
For example, suppose there are two planets in a system and both planets are observed with $b\approx0$.  Without additional information, the resulting configuration could be due to both planets having low relative inclinations and the observer's line of sight happening to be along the mutual orbital plane. 
It could also be the case that the planets have very large relative orbital inclinations $I$, but one or both nodes happen to be nearly aligned with the observer's line of sight.  
For arbitrary combinations of $b$'s (or equivalently $i$'s), a wide range of $\vec{I}$ could be possible.

Fortunately, this degeneracy between $i$ and $I$ can be broken in principle through TDVs, revealing or at least constraining the 3D plane of the planets \citep[e.g.,][]{agol_carter_2018}.  
Similar to transit timing variations \citep[TTVs; ][]{ford_etal_2012}, perturbations by other planets or even the stellar potential (and for very hot Jupiters the planet's potential \citep{ragozzine_wolf_2009}) will cause changes to the planetary orbits and the chord lengths of corresponding transits. 
The magnitude and direction of the resulting TDVs can in turn be used to model the actual inclination vectors of the planets for a given system.  
Should it prove not to be possible to match a system's TDVs with orbital solutions, then the discrepancy could point to additional perturbations, such as non-transiting planets.  

To date, observations of TDVs are limited among the confirmed planet sample, but the number of systems with measured TDVs is expanding.  
Examples include Kepler-13 \citep{szabo_etal_2012,mazeh_etal_2013}, Kepler-88 \citep{nesvorny_etal_2013}, and Kepler-108 \citep{mills_fabrycky_2017}.
Variations in Kepler-13 b are consistent with evolution due to the star's stellar oblateness, while the TDVs in Kepler-88 are due to eccentricity variations among two planets in a 2:1 resonance.  
Kepler-108 exhibits TDVs consistent with changes in the impact parameter resulting from planet-planet interactions.  
Photodynamical modelling of Kepler-9c \citep{freudenthal_etal_2018} revealed that the planet duration should shorten and the planet should eventually stop transiting in the near future. 
An extensive search of Kepler Objects of Interest (KOIs)  by \citet{kane_etal_2019} also found roughly 100 TDV candidates. 
 Finally, K2-146c was recently observed to come into transit \citep{hamann_etal_2019}, demonstrating that planets can indeed re-transit and presumably de-transit.
These results demonstrate that TDVs are measurable and contain a wealth of possibilities for examining dynamical interactions within a planetary system.

\begin{figure}
\includegraphics[width=4in,angle=-90]{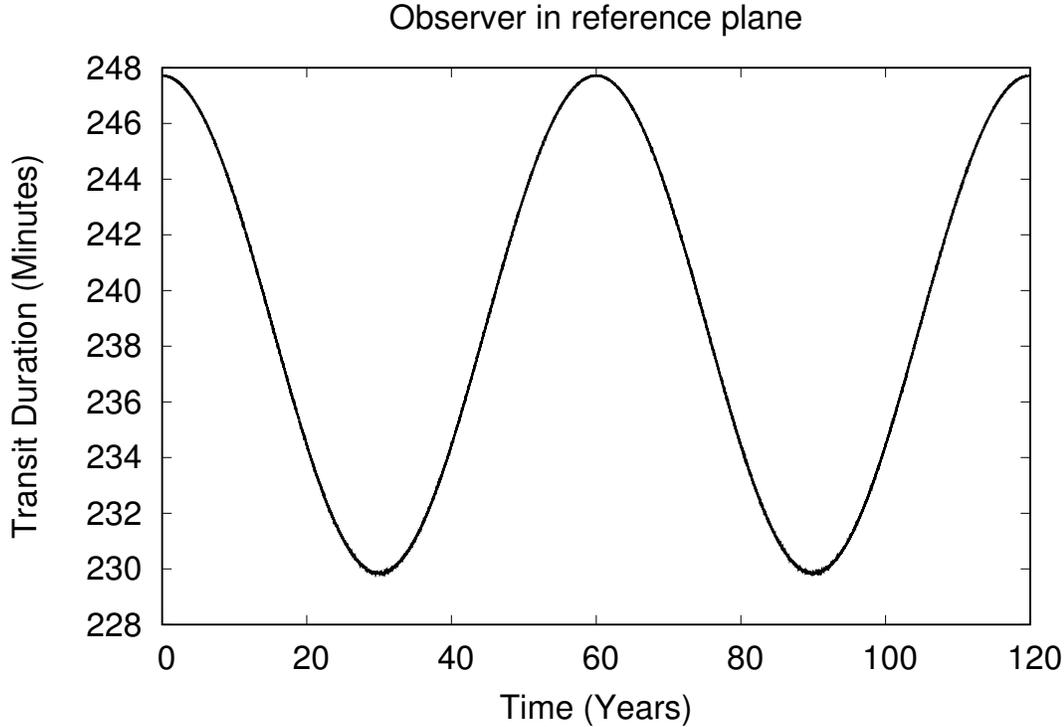}
\caption{Example of a TDV due only to  precession of $\Omega$, which is assumed in this toy model to be $\dot\Omega = 3^\circ~\rm yr^{-1}$. 
The star's radius is 1$R_{\odot}$, and the circular orbit of the planet has a semi-major axis of $a=0.1$ au and an inclination $I=1^\circ$.  
The observer is assumed to lie along the reference plane.  
Precession alone can cause significant chord changes for moderate inclinations, leading to transit duration variations $>5$ min on decade timescales.} \label{fig:tdv_omega}
\end{figure}

Secular perturbations among the planets in a STIP will be a major driver of long-term TDVs by varying $\vec{I}$.
We stress that precession of $\Omega$ alone can lead to changes in a transit's chord length, i.e., the magnitude of the inclination need not change to produce an observable signature.  
This is highlighted in Figure \ref{fig:tdv_omega}, in which a planet on a circular orbit with semi-major axis $a=0.1$ au orbits a solar-sized star.  
The planet's size is not included in this toy model, and the observer is assumed to be in the reference plane.  
A precession rate of $3^\circ~\rm~yr^{-1}$ is assumed, leading to large TDVs on decade timescales.  
Whether real systems commonly have such observable precession rates is one of the focuses of this work.

A recent study that examined potential TDVs among multiplanet systems \citep{becker_adams_2016} used Lagrange-Laplace secular theory to determine how the orbital inclinations of planets in STIPs would vary with time, and used the results to calculate the expected TDVs for known systems.  The work also assessed whether transiting planets in STIPs might move away from transiting configurations (i.e., de-transit) due to self-interactions. 
Their results focused on changes to the magnitude of orbital inclinations ($I$), but did not explicitly include the contribution from variations in $\Omega$.
In particular, they found that de-transits should be rare.

In this work, we re-examine TDV signatures and the potential for systems to de-transit by taking into account changes in the full $\vec{I}$.
We begin in section 2 by  using Laplace-Lagrange secular theory to evaluate the nodal precession frequencies  of known systems, which is used as an indication of the TDV timescale.   
Systems with rapid precession frequencies and large impact parameters together can be used to select systems that should show strong TDVs.
We then directly simulate several known STIPs in section 3 to produce synthetic TDVs, determine potential variations on decadal timescales, and assess those systems for potential evolution between transiting and non-transiting states. 
The results of those simulations are presented in section 4 followed by a discussion in section 5.  The main findings are summarized in section 6.

\section{Indicators of Noticeable TDVs}

\subsection{Secular Inclination Frequencies}\label{sec:secular}

We use Laplace-Lagrange secular theory to determine the eigenfrequency structure for 118 STIPs.  
In short, the secular theory assumes that short-period gravitational perturbations average out on long timescales, which is valid provided that the planets are not strongly interacting in, for example, mean-motion resonances (MMRs) or near-MMRs.  
The remaining long-period perturbations affect the eccentricity and inclination vectors, but not the semi-major axes. 
Secular theory is second order in $e$ and $I$, and is first-order in mass. 
Eccentricity and inclination are decoupled at this order.  
If the planets are near an MMR, then the system could still be described by secular theory, but additional modes may be present and the actual modes may be shifted compared with the secular eigenfrequencies. 
Nonetheless, for strong planet-planet interactions, the actual nodal eigenfrequencies (see below) can show excellent correspondence to the Laplace-Lagrange theory even if the apsidal eigenfrequencies do not \citep[e.g., ][]{malhotra_etal_1999,granados_boley_2018}. 
We refer the reader to \cite{murray_dermott_1999} for a thorough overview; several key relations are listed for clarity.  

For each planet $j$, the inclination vector (with magnitude $I_j$ and direction $\Omega_j$) can be described by the rectangular coordinates
\begin{eqnarray}
q_j &=&  I_j \cos \Omega_j\rm~and \\
p_j &=&  I_j \sin \Omega_j.
\end{eqnarray}
The evolution of $q,p$ is governed by coupled, linear, first-order differential equations. 
A similar set of equations can be written to describe the eccentricity vector (not shown).  
Thus, the long-term evolution of the planetary orbits can be reduced to a series summation of sines and cosines, explicitly dependent on time and running over the eigenfrequencies of the system:
\begin{eqnarray}
q_j &=& \sum_{i=1}^{N-1} \mathcal{I}_i U_{ij} \cos(f_i t + \gamma_i),~\rm and \\
p_j &= & \sum_{i=1}^{N-1} \mathcal{I}_i U_{ij} \sin(f_i t + \gamma_i).
\end{eqnarray}
For $N$ planets, there are $N$ apsidal (eccentricity) eigenfrequencies and $N-1$ non-zero nodal (inclination) eigenfrequencies, as one degree of freedom is lost in defining the mean plane of the system.  
The eigenfrequencies $f_i$ depend only on the system masses and the planets' orbital semi-major axes, as do the eigenvectors $U_{ij}$ for a specific mode $i$ on planet $j$.
Secular theory itself is unscaled. The mode amplitudes ($\mathcal{I}_i$) and phases ($\gamma_i$) must be determined from boundary conditions -- the orbital  inclination $I$ and node $\Omega$ at a given time.

Because we expect the inclinations of most systems to be reasonably described by Laplace-Lagrange secular theory, we can use the inclination eigenfrequencies as a preliminary ranking for systems that should be targeted for follow up.
To that end, we use masses and semi-major axes from the NASA exoplanets archive\footnote{https://exoplanetarchive.ipac.caltech.edu/, last accessed 28 December 2018} to determine the inclination eigenfrequencies of various systems. 
We only consider systems that have three or more planets.  We further require that all of the known planets are currently transiting and have well-defined semi-major axes or periods.
The archive's ``default'' values are selected when possible.
If masses are available from the archive, then those are used directly. 
Otherwise, the \cite{lissauer_etal_2011} mass-radius relation is used to estimate planetary masses with the understanding that this is only approximate.  
Specifically, we used $M/M_\oplus \approx (R/R_\oplus)^q$ with $q=2.01$. This is slightly different from the Lissauer et al.~relation that uses $q=2.06$.  This was an unintentional change discovered after the analysis had been completed, but because the overall effect is small and well within the uncertainties, we continued with the smaller $q$.  

As an example, Table~\ref{tbl:secfreqtbl} lists all nodal eigenfrequencies for select exoplanet systems.  The systems are arranged by multiplicity and then Kepler number, while the eigenfrequencies are sorted by rate. Multiple frequencies exceed one degree per year, suggesting that TDVs in these systems could be observable on relatively short timescales.   We note that the uncertainty in mass measurements will give rise to a comparable uncertainty in the eigenfrequencies.  For well-constrained systems, we expect our estimates to be accurate to tens of percent, while for poorly constrained systems, the eigenfrequencies should still yield an order of magntidue estimate, on average.   

\begin{table}
\caption{Secular eigenfrequencies from Laplace-Lagrange secular theory for select exoplanet systems.  The results are sensitive to the planetary masses and orbital semi-major axes, so frequencies are only shown to the nearest tenth of a degree per year.  Values of zero are not intended to mean exactly zero; rather, the magnitude of the value is less than $0.05^\circ~yr^{-1}$. }\label{tbl:secfreqtbl}

\begin{tabular}{lclllllll}
 & & \multicolumn{7}{l}{Secular Frequencies (deg per yr)} \\
System & $N_{planets}$ & $f_1$ & $f_2$ & $f_3$ & $f_4$ & $f_5$ & $f_6$ & $f_7$ \\
\hline\hline
KOI351 & 8 &   -0.7 &   -0.7 &   -0.4 &   -0.3 &   -0.2 &   -0.1 &   -0.0 \\ 
TRAPPIST-1 & 7 &   -9.5 &   -5.3 &   -4.9 &   -2.8 &   -1.4 &   -0.6 \\ 
Kepler-80 & 6 &   -4.2 &   -2.9 &   -1.5 &   -0.8 &   -0.4 \\ 
Kepler-11 & 6 &   -1.2 &   -0.8 &   -0.3 &   -0.2 &   -0.0 \\ 
HIP 41378 & 5 &   -1.5 &   -0.2 &   -0.1 &   -0.0 \\ 
K2-138 & 5 &   -4.5 &   -3.4 &   -2.1 &   -1.1 \\ 
Kepler-32 & 5 &   -1.2 &   -0.9 &   -0.2 &   -0.2 \\ 
Kepler-33 & 5 &   -2.6 &   -1.5 &   -0.9 &   -0.4 \\ 
Kepler-55 & 5 &   -0.6 &   -0.4 &   -0.3 &   -0.1 \\ 
Kepler-62 & 5 &   -0.4 &   -0.1 &   -0.0 &   -0.0 \\ 
Kepler-84 & 5 &   -1.0 &   -0.3 &   -0.2 &   -0.1 \\ 
Kepler-102 & 5 &   -0.7 &   -0.5 &   -0.3 &   -0.2 \\ 
Kepler-122 & 5 &   -1.3 &   -1.1 &   -0.4 &   -0.2 \\ 
Kepler-154 & 5 &   -0.5 &   -0.4 &   -0.1 &   -0.1 \\ 
Kepler-169 & 5 &   -1.1 &   -0.4 &   -0.2 &   -0.0 \\ 
Kepler-186 & 5 &   -0.8 &   -0.2 &   -0.1 &   -0.0 \\ 
Kepler-238 & 5 &   -4.7 &   -1.1 &   -0.8 &   -0.4 \\ 
Kepler-292 & 5 &   -1.8 &   -1.0 &   -0.5 &   -0.3 \\ 
Kepler-296 & 5 &   -0.7 &   -0.4 &   -0.2 &   -0.1 \\ 
Kepler-444 & 5 &   -0.5 &   -0.4 &   -0.3 &   -0.1 \\ 
\end{tabular}
\end{table}

\subsection{Considering the Impact Parameter}\label{sec:imparam}

Even fast nodal frequencies may have little observational impact if the magnitude of the inclination is small. 
Because we do not know the actual orbital inclinations, we use the impact parameters as a general indicator.  
This should be taken with caution.
The impact parameter only tells us the inclination to our line of sight, and it is possible for two planets with the same inferred inclination to have non-zero mutual inclinations.
It is also possible to have a system with large impact parameters, but small orbital inclinations if the system as a whole is inclined to the observer.  
In this case, one would expect the impact parameters for planets in a system to increase with increasing semi-major axis.

With these caveats in mind, impact parameters, while having large uncertainties, provide additional information for assessing whether a particular system may exhibit strong TDVs. 

\subsection{Nodal Frequency Ranking} \label{sec:secevol}

\begin{figure}
\centering
\includegraphics[width=0.5\textwidth]{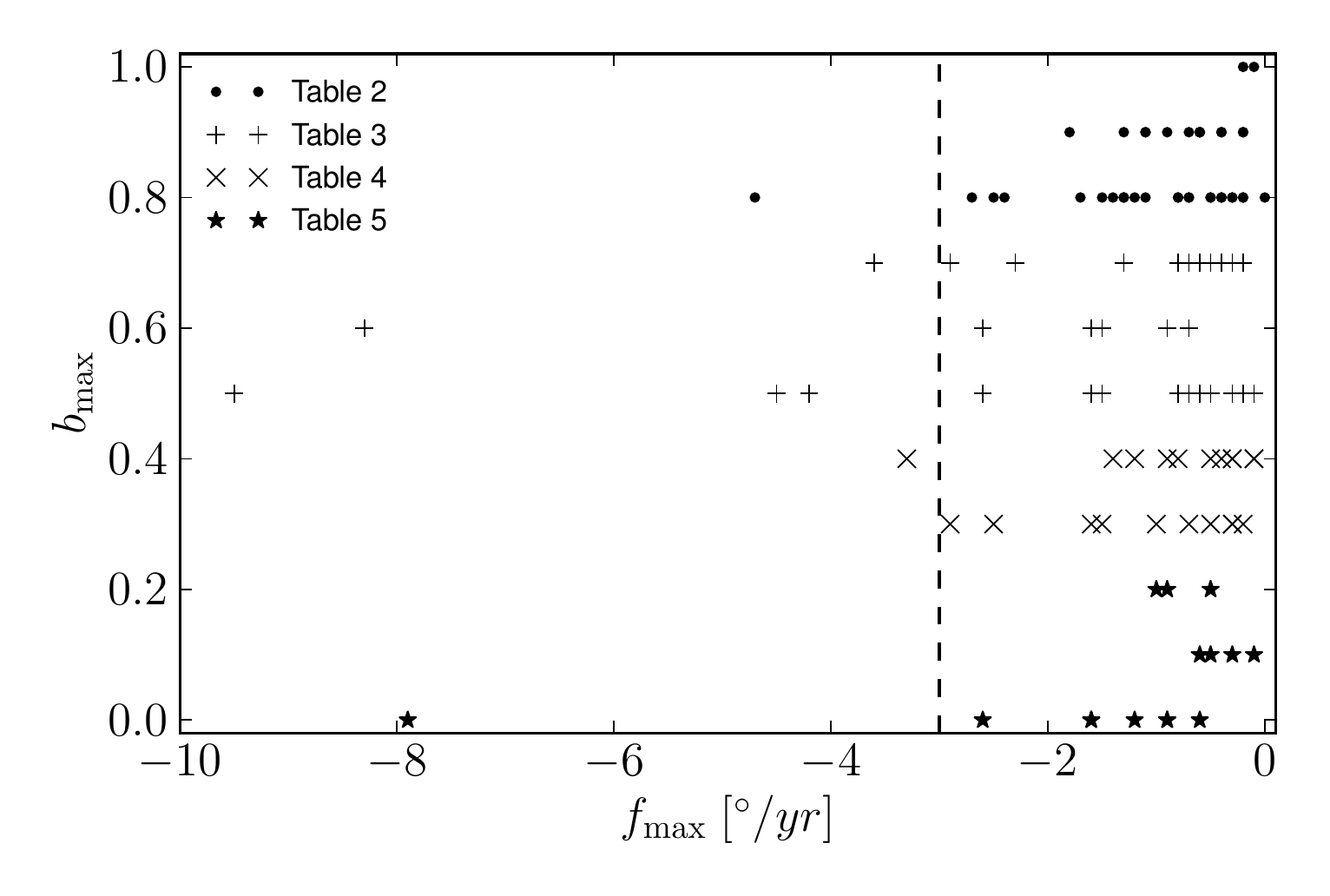}
\caption{Fastest secular eigenfrequency, $f_{\rm max}$, vs.~maximum transit impact parameter, $b_{\rm max}$, of the planetary systems listed in Tables \ref{tbl:rank_last} to \ref{tbl:rank_first}. The different symbols represent the separation of impact parameter in quartiles. The vertical dashed line defines those systems with the highest secular eigenfrequencies, as listed in Table \ref{tbl:rank_frequency}.} \label{fig:rank_frequency}
\end{figure}

Tables~\ref{tbl:rank_last} through \ref{tbl:rank_first} rank 118 STIPs from the exoplanet archive according to the following: systems are first separated into quartiles by the maximum transit impact parameter in each system. Within each quartile, the systems are then sorted by the largest nodal eigenfrequency.  Entries listed with a maximum frequency of zero are not formally zero, but have a magnitude less than $0.05^\circ~{\rm yr}^{-1}$.  These tables are summarized in Figure \ref{fig:rank_frequency}.

The motivation for using quartiles is to highlight which STIPs may be the most promising for future followup according to both $b$ and nodal precession.
Because the uncertainties in the impact parameters may make such groupings misleading, we also show in Table \ref{tbl:rank_frequency} the top ten systems with the fastest nodal eigenfrequencies. 
The systems all lie to the left of the vertical line in Figure \ref{fig:rank_frequency}, which have a maximum secular eigenfrequency faster than $-3.0^\circ~{\rm yr}^{-1}$.

While high multiplicity systems tend to be in the upper half of any given quartile, they do not in general have the fastest eigenfrequencies.   Thus, the number of planets in a system is not necessarily a good predictor of large TDVs, at least based on secular analysis.  

 For systems  in Table \ref{tbl:rank_frequency}, the planets could exhibit TDVs from precession alone (e.g., Fig.~\ref{fig:tdv_omega}) due to the high eigenfrequencies.  This does not consider the magnitude of the inclination vector, which will have a large effect on the actual TDV signal. This will particularly be highlighted with Kepler-445 using direct N-body simulation, discussed next.

\begin{table}
\caption{Systems ranked by fastest secular eigenfrequency for maximum impact parameters $0.75 \le b < 1$.
  }\label{tbl:rank_last}
\begin{tabular}{lcll}
System & $N_{planets}$ & $f_{max}$ (deg per yr) & $b_{max}$  \\
\hline\hline
Kepler-238 & 5 & -4.7 & 0.8 \\ 
Kepler-304 & 4 & -2.7 & 0.8 \\ 
  Kepler-46 &3 &   -2.5 & 0.8  \\
    K2-37  &3  &   -2.4 & 0.8  \\
Kepler-292 & 5 & -1.8 & 0.9 \\ 
Kepler-26 & 4 & -1.7 & 0.8 \\ 
Kepler-256 & 4 & -1.5 & 0.8 \\ 
Kepler-224 & 4 & -1.4 & 0.8 \\ 
Kepler-306 & 4 & -1.3 & 0.9 \\
 Kepler-18  & 3   &   -1.3 & 0.8   \\
Kepler-122 & 5 & -1.3 & 0.8 \\ 
Kepler-11 & 6 & -1.2 & 0.8 \\ 
  Kepler-58 &3 &   -1.1 & 0.9  \\
   GJ 9827 &  3   &  -1.1 & 0.9  \\
   Kepler-169 & 5 & -1.1 & 0.8 \\ 
     Kepler-359 & 3 &   -0.9 & 0.9  \\
Kepler-172 & 4 & -0.8 & 0.8 \\
  K2-32 & 3  &  -0.8  & 0.8 \\
   Kepler-326 &3 &   -0.7 & 0.9 \\
  Kepler-250 &3 &   -0.7 & 0.8  \\
Kepler-338 & 4 & -0.7 & 0.8 \\ 
     Kepler-138 &3&   -0.6 &0.9  \\
  Kepler-142 &3 &   -0.6 & 0.9  \\
     Kepler-272 &3 &   -0.6 & 0.9  \\
      Kepler-276 &3 &   -0.6 & 0.9  \\
Kepler-215 & 4 & -0.5 & 0.8 \\ 
Kepler-444 & 5 & -0.5 & 0.8 \\ 
Kepler-299 & 4 & -0.4 & 0.9 \\ 
   Kepler-351 &3 &   -0.4 & 0.9  \\
 Kepler-53 &3 &   -0.4 & 0.8  \\
  Kepler-104 &3 &   -0.4 & 0.8  \\
Kepler-342 & 4 & -0.3 & 0.8 \\
 Kepler-52 &3 &   -0.3 & 0.8  \\
 Kepler-79 & 4 & -0.2 & 1.0 \\
  Kepler-310 &3 &   -0.2 & 0.9  \\
 Kepler-357 &3 &   -0.2 & 0.9  \\
    Kepler-100 &3 &   -0.2 & 0.8   \\
  Kepler-298 &3 &   -0.2 & 0.8  \\
 Kepler-51 &3 &   -0.1 & 1  \\
  Kepler-130 &3 &   -0.0 & 0.8  \\
\end{tabular}
\end{table}

\begin{table}
\caption{Systems ranked by fastest secular eigenfrequency for impact parameters $0.5 \le b < 0.75$}\label{tbl:rank_third}
\begin{tabular}{lcll}
System & $N_{planets}$ & $f_{max}$ (deg per yr) & $b_{max}$  \\
\hline\hline
TRAPPIST-1 & 7 & -9.5 & 0.5 \\ 
    K2-19 & 3   &   -8.3 & 0.6 \\
K2-138 & 5 & -4.5 & 0.5 \\ 
Kepler-80 & 6 & -4.2 & 0.5 \\ 
 Kepler-42 &3 &   -3.6 & 0.7 \\
  Kepler-446 &3 &   -2.9 & 0.7  \\
   Kepler-305 &3 &   -2.6 & 0.6  \\
Kepler-33 & 5 & -2.6 & 0.5 \\ 
       Kepler-114 &3 &   -2.3 & 0.7  \\
Kepler-221 & 4 & -1.6 & 0.6 \\
    Kepler-203 &3 &   -1.6 & 0.5 \\
HIP 41378 & 5 & -1.5 & 0.6 \\ 
 Kepler-65 &3 &   -1.5 & 0.5  \\
  Kepler-279 &3 &   -1.3 & 0.7  \\
   Kepler-319 &3 &  -0.9  & 0.6 \\
 Kepler-339 &3 &   -0.9 & 0.6  \\
 Kepler-275 &3 &   -0.8 & 0.7  \\
  Kepler-374 &3 &   -0.8 & 0.7 \\
   Kepler-267 &3 &   -0.8 & 0.5  \\
Kepler-296 & 5 & -0.7 & 0.7 \\ 
Kepler-102 & 5 & -0.7 & 0.6 \\ 
KOI-351 & 8 & -0.7 & 0.5 \\ 
Kepler-55 & 5 & -0.6 & 0.7 \\ 
    Kepler-81 &3 &   -0.6 & 0.7  \\
Kepler-167 & 4 & -0.6 & 0.5 \\ 
   K2-136 & 3  & -0.6 & 0.5  \\
Kepler-154 & 5 & -0.5 & 0.7 \\ 
Kepler-341 & 4 & -0.5 & 0.7 \\ 
  Kepler-31 &3&    -0.5 & 0.5   \\
           Kepler-235 & 4 & -0.4 & 0.7 \\ 
         Kepler-363 &3 &   -0.4 & 0.7  \\
Kepler-106 & 4 & -0.3 & 0.7 \\ 
Kepler-327 &3 &   -0.3 & 0.7  \\
Kepler-354 &3 &   -0.3 & 0.7  \\
 Kepler-164 &3 &   -0.3 & 0.5  \\
 Kepler-331 &3 &   -0.3 & 0.5  \\
    K2-155 & 3 & -0.3 & 0.5    \\
    Kepler-265 & 4 & -0.2 & 0.7 \\
      K2-3 & 3  &    -0.2  & 0.7 \\
     K2-183 & 3  &   -0.2 & 0.5   \\
 Kepler-336 &3 &   -0.2 & 0.5  \\
    Kepler-332 &3 &   -0.1 & 0.5  \\
\end{tabular}
\end{table}

\begin{table}
\caption{Systems ranked by fastest secular eigenfrequency for impact parameters $0.25 \le b < 0.5$}\label{tbl:rank_second}
\begin{tabular}{lcll}
System & $N_{planets}$ & $f_{max}$ (deg per yr) & $b_{max}$  \\
\hline\hline
Kepler-30 &3&   -3.3 & 0.4  \\
 Kepler-226 &3 &   -2.9 & 0.3  \\
Kepler-49 & 4 & -2.5 & 0.3 \\ 
Kepler-23 &3&   -1.6 & 0.3  \\
Kepler-24 & 4 & -1.5 & 0.3 \\ 
KOI-94 & 4 & -1.4 & 0.4 \\ 
 Kepler-350 &3 &   -1.2 & 0.4  \\
Kepler-84 & 5 & -1.0 & 0.3 \\ 
 Kepler-83 &3 &   -0.9 & 0.4  \\
Kepler-186 & 5 & -0.8 & 0.4 \\ 
 Kepler-255 &3 &   -0.7 & 0.3\\
Kepler-208 & 4 & -0.5 & 0.3 \\ 
Kepler-289 &3 &   -0.5 & 0.4  \\
Kepler-62 & 5 & -0.4 & 0.4 \\ 
Kepler-206 &3 &   -0.3 & 0.4  \\
Kepler-325 &3 &   -0.3 & 0.4  \\
Kepler-251 & 4 & -0.3 & 0.3 \\ 
   Kepler-301 &3 &   -0.3 & 0.3  \\
      Kepler-372 &3 &   -0.2 & 0.3  \\
       Kepler-198 &3 &   -0.1 & 0.4  \\
    Kepler-288 &3 &   -0.1 & 0.4  \\

\end{tabular}
\end{table}

\begin{table}
\caption{Systems ranked by fastest secular eigenfrequency for impact parameters $b < 0.25$}\label{tbl:rank_first}
\begin{tabular}{lcll}
System & $N_{planets}$ & $f_{max}$ (deg per yr) & $b_{max}$  \\
\hline\hline
Kepler-445 &3 & -7.9 & 0  \\
Kepler-223 & 4 & -2.6 & 0.0 \\ 
Kepler-1542 & 4 & -1.6 & 0.0 \\ 
K2-072 & 4 & -1.6 & 0.0 \\ 
Kepler-32 & 5 & -1.2 & 0.0 \\ 
 Kepler-54 &3 &   -1.0 & 0.2  \\
 Kepler-85 & 4 & -0.9 & 0.2 \\ 
Kepler-1388 & 4 & -0.9 & 0.0 \\ 
Kepler-758 & 4 & -0.9 & 0.0 \\ 
Kepler-191 &3 &   -0.6 & 0.1  \\
K2-187 & 4 & -0.6 & 0.0 \\
Kepler-245 & 4 & -0.5 & 0.2 \\ 
Kepler-82 & 4 & -0.5 & 0.1 \\ 
Kepler-282 & 4 & -0.3 & 0.1 \\ 
  Kepler-334 &3 &   -0.1 & 0.1  \\
\end{tabular}
\end{table}

\begin{table}
\caption{The ten systems in our sample with the highest nodal eigenfrequencies, ranked from fastest to slowest.
  }\label{tbl:rank_frequency}
\begin{tabular}{lcll}
System & $N_{planets}$ & $f_{max}$ (deg per yr) & $b_{max}$  \\
\hline\hline
TRAPPIST-1 & 7 & -9.5 & 0.5 \\ 
    K2-19 & 3   &   -8.3 & 0.6 \\
    Kepler-445 &3 & -7.9 & 0  \\
Kepler-238 & 5 & -4.7 & 0.8 \\ 
K2-138 & 5 & -4.5 & 0.5 \\ 
Kepler-80 & 6 & -4.2 & 0.5 \\ 
 Kepler-42 &3 &   -3.6 & 0.7 \\
Kepler-30 &3&   -3.3 & 0.4  \\
  Kepler-446 &3 &   -2.9 & 0.7  \\
 Kepler-226 &3 &   -2.9 & 0.3  \\
\end{tabular}
\end{table}

\section{N-body Simulations}\label{sec:ICs}

Ranking systems based on an eigenfrequency analysis under Laplace-Lagrange theory only gives a sense for which systems could show large TDVs.  
It does not give a general understanding of the range TDVs that are possible for a given system, nor does it elucidate the fraction of systems that could have planets going in and out of transit.  
We thus select several systems and run direct N-body simulations to capture the evolution of planetary transits for several assumptions regarding the distribution of inclination vectors.  
All realizations are constrained by transit observations.
The systems we choose to explore are Kepler-11, TRAPPIST-1, K2-138, Kepler-445, and Kepler-334.  
The first two are chosen because of their general interest within the exoplanet community.  K2-138\footnote{An additional planet may exist in this system \citep{christiansen_etal_2018}.  For this study, we have focused solely on the confirmed planets.} is an additional system that has a large maximum impact parameter and fast nodal eigenfrequencies.  Kepler-445 has a nearly zero maximum impact parameter, but a high maximum eigenfrequency.  Thus, depending on the orientation of the system relative to the observer, the realizations could exhibit large to negligible TDVs.  
The final system, Kepler-334, has a very low maximum eigenfrequency.  Realizations of this system should show negligible TDVs, regardless of the system orientation to the observer.

\subsection{Initial Conditions}

Because we want to model a given system as accurately as  practicable, we use a combination of the NASA exoplanet archive along with updated observational constraints, if available, to determine the initial configuration of the system. 
Table \ref{tbl:icsims} lists the masses, semi-major axes, eccentricities (if known), and current mid-transit times used in the simulations.  This does assume that there are no significant non-transiting planets in the system.  If the eccentricity is unknown for a planet, it is drawn from a Rayleigh distribution with $\sigma_e=0.0025$, chosen to be low but non-trivial. We note that this is lower than that reported by \cite{vaneylen_2015}, who found that the eccentricities for planets within a subset of Kepler multiplanet systems is roughly Rayleigh distributed with $\sigma_e=0.049\pm0.013$. Larger eccentricities would lead to more unstable systems among our initial conditions, so our choice of a lower $\sigma_e$ helps to bias the initial conditions toward stable realizations. 

All simulations are referenced to an initial time of ${\rm JD}=2458224.50000$.
The planets are given true anomalies that are consistent with the published transit midpoints.
The longitudes of pericentre are drawn from a uniform random distribution between $0^\circ$ and $360^\circ$, except the first five planets of Kepler-11, which have longitudes constrained with the eccentricities.  
Orbital inclinations and the longitudes of ascending node $\Omega$ are randomly determined, constrained by the system orientation and the published impact parameters.
Due to degeneracies between orbital inclinations and the locations of the nodes, we use several different sampling distributions to explore the range of allowed inclination vectors, as described next.

\begin{table}
\caption{ICs for the N-body simulations.  
Values are from \url{https://exoplanetarchive.ipac.caltech.edu/index.html} when available. Specific references are as follows:  \cite{delrez_etal_2018} and \cite{grimm_etal_2018} for Trappist-1; \cite{lissauer_etal_k11_2013} for Kepler-11; \cite{christiansen_etal_2018} for K2-138; \cite{muirhead_etal_2015} for Kepler-445; and \cite{rowe_etal_2014} for K334. 
Among the Kepler-11 realizations, the longitudes of pericentre are determined for all but the outermost planet, and as such, we set $\varpi=135^\circ,~141^\circ,~56^\circ,~138^\circ,\rm~and~66^\circ$ for the first five planets and set $\varpi$ to be random for the last planet. The longitudes are randomly determined between 0 and $2\pi$ for the rest of the systems.    \label{tbl:icsims}}
\begin{tabular}{lllllll}
Planet &  $a$ [au]  & Radius [$\rm R_\oplus$] &   Mass [$\rm M_\oplus$] & $e$ & b & Transit mid [JD] \\
\hline\hline
Trappist-1 & & $0.121\pm 0.003$\tablenotemark{a}  & $0.089\pm0.006$\tablenotemark{b} & & & \\
Trappist-1 b & $0.0115^{+0.00028}_{-0.00025}$ & $1.127\pm 0.028$ & $1.017\pm 0.15$ & - & $0.157\pm 0.075$ & $2457322.51654\pm 0.00010$\\
Trappist-1 c & $0.01576^{+0.00038}_{-0.00034}$ & $1.1\pm 0.028$ & $1.156\pm 0.14$ & - & $0.148\pm 0.088$ & $2457282.80879\pm 0.00018$\\
Trappist-1 d & $0.02219^{+0.00053}_{-0.00048}$ & $0.788\pm 0.020$ & $0.297\pm 0.037$ & - & $0.08^{+0.10}_{-0.06}$ & $2457670.14227\pm 0.00026$ \\
Trappist-1 e & $0.02916^{+0.00070}_{-0.00063}$ & $0.915\pm 0.025$ & $0.772\pm 0.077$ & - & $0.240^{+0.056}_{-0.047}$ & $2457660.37910\pm 0.00040$ \\
Trappist-1 f & $0.03836^{+0.00092}_{-0.00084}$ & $1.052\pm 0.026$ & $0.934\pm0.79$ & - & $0.337^{+0.040}_{-0.029}$ &  $2457671.39470\pm 0.00022$ \\
Trappist-1 g & $0.0467\pm 0.0011$ & $1.154\pm 0.029$ & $1.148\pm 0.097$ & - & $0.406^{+0.031}_{-0.025}$ &  $2457665.35084\pm 0.00020$\\
Trappist-1 h & $0.0617^{+0.0015}_{-0.0013}$ & $0.777\pm 0.025$ & $0.331\pm 0.053$ & - & $0.392^{+0.039}_{-0.043}$ &  $2457662.55467\pm 0.00054$ \\
\hline\hline 

Kepler-11 & & $0.961\pm 0.025$\tablenotemark{a} & $1.065^{+0.0017}_{-0.022}$\tablenotemark{b} & & & \\
Kepler-11 b &  $0.091\pm 0.001$ & $1.8^{+0.03}_{-0.05}$  & $1.9^{+1.4}_{-1.0}$  & $0.045^{+0.068}_{-0.042}$ & $0.116^{+0.053}_{-0.116}$ & $2455589.7378^{+0.0026}_{-0.0047}$  \\
Kepler-11 c &  $0.107\pm 0.001$ & $2.87^{+0.05}_{-0.06}$ & $2.9^{+2.9}_{-1.6}$  & $0.026^{+0.063}_{-0.026}$ & $0.156^{+0.059}_{-0.156}$ & $2455583.3494^{+0.0014}_{-0.0019}$  \\
Kepler-11 d &  $0.155\pm 0.001$ & $3.12^{+0.06}_{-0.07}$ & $7.3^{+0.8}_{-1.5}$  & $0.004^{+0.007}_{-0.002}$ & $0.181^{+0.074}_{-0.084}$ & $2455594.0069^{+0.0022}_{-0.0014}$  \\
Kepler-11 e &  $0.195\pm 0.002$ & $4.19^{+0.07}_{-0.09}$ & $8.0^{+1.5}_{-2.1}$  & $0.012\pm 0.006$ & $0.763\pm 0.008$ & $2455595.0755^{+0.0015}_{-0.0009}$  \\
Kepler-11 f &  $0.250\pm 0.002$ & $2.49^{+0.04}_{-0.07}$ & $2.0^{+0.8}_{-0.9}$  & $0.013^{+0.011}_{-0.009}$ & $0.463^{+0.030}_{-0.032}$ & $2455618.2710^{+0.0041}_{-0.0038}$  \\
Kepler-11 g &  $0.466\pm 0.004$ & $3.33^{+0.06}_{-0.08}$ & 11\tablenotemark{c} & 0.013\tablenotemark{d} & $0.217^{+0.092}_{-0.087}$ & $2455593.8021^{+0.0030}_{-0.0021}$  \\
\hline\hline

K2-138 & & $0.93\pm0.06$\tablenotemark{a} & $0.86\pm0.08$\tablenotemark{b} & & & \\
K2-138 b & $0.03380\pm 0.00024$ & $1.57^{+0.28}_{-0.17}$ & 2.5\tablenotemark{c} & - & $0.50^{0.33}_{-0.34}$ & $2457773.317^{+0.0037}_{-0.0038}$ \\
K2-138 c & $0.04454\pm 0.00032$ & $2.52^{+0.34}_{-0.16}$ & 6.4\tablenotemark{c} & - & $0.47\pm 0.32$ & $2457740.3223^{+0.0025}_{-0.027}$ \\
K2-138 d & $0.05883\pm 0.00042$ & $2.66^{+0.39}_{-0.18}$ & 7.1\tablenotemark{c} & - & $0.47^{+0.31}_{-0.32}$ &      $2457743.1607^{+0.0036}_{-0.0037}$ \\
K2-138 e & $0.07807\pm 0.00056$ & $3.29^{+0.35}_{-0.18}$ & 11\tablenotemark{c} & - & $0.44^{0.31}_{-0.30}$ &      $2457740.6451^{+0.0020}_{-0.0021}$ \\
K2-138 f & $0.1043^{+0.00074}_{-0.00075}$  & $2.81^{+0.36}_{-0.19}$ & 8.0\tablenotemark{c} & - & $0.48^{+0.30}_{-0.33}$ &     $2457738.7019^{+0.0033}_{-0.0035}$ \\
\hline\hline

Kepler-445  & & $0.21\pm0.03$\tablenotemark{a}   & $0.16\pm 0.04$\tablenotemark{b} &  & & \\
Kepler-445 b & 0.0229\tablenotemark{e} & $1.58\pm 0.23$ & 2.5\tablenotemark{c} & - & 0.01\tablenotemark{f} & $2454966.1194\pm 0.0033$\\
Kepler-445 c & 0.0318\tablenotemark{e} & $2.51\pm 0.36$ & 6.4\tablenotemark{c} & - & 0.01\tablenotemark{f} & $2454966.6408\pm 0.0019$\\
Kepler-445 d & 0.0448\tablenotemark{e} & $1.25\pm 0.19$ & 1.6\tablenotemark{c} & - & 0.01\tablenotemark{f} &  $2454836.751\pm 0.05$ \\
\hline\hline

Kepler-334 & & $1$\tablenotemark{a,g}   &  $1$ \tablenotemark{b} & & & \\
Kepler-334 b & 0.061\tablenotemark{e} & $1.12\pm 0.21$   &  1.3 \tablenotemark{c} & - &  $0.11^{+0.2}_{-0.11}$ & $2454964.49467\pm 0.0021$\\
Kepler-334 c & 0.107\tablenotemark{e} &  $1.43\pm 0.26$  &  2.1 \tablenotemark{c}  & -  & $0.03^{+0.2}_{-0.03}$ &  $2454966.95790\pm0.0026$\\
Kepler-334  d &  0.168\tablenotemark{e} &  $1.41\pm0.26$  &  2.0  \tablenotemark{c} &  -   & $0.07^{+0.2}_{-0.07}$ & $2454978.56945\pm0.0035$  \\\hline

\end{tabular}
\tablenotetext{a} {This is the star's radius given in $\rm R_{\odot}$.}
\tablenotetext{b} {This is the star's mass given in $\rm M_{\odot}$.}
\tablenotetext{c}{Masses inferred using the \cite{lissauer_etal_2011} radius-mass relation, but using $M\approx \left(\frac{R}{R_\oplus}\right)^q \rm M_\oplus $ with $q=2.01$ instead of $q=2.06$. See text.}
\tablenotetext{d}{The value is well below the upper limit of 0.1, and is taken to be the same as Kepler-11 f.}
\tablenotetext{e}{The semi-major axes are determined from the published periods using the constrained stellar mass. }
\tablenotetext{f}{Published value is nearly zero with very large uncertainties. Using a value that is small and well-behaved in the IC generator.  }
\tablenotetext{g}{The stellar mass is assumed to be $1~\rm M_\odot$}
\end{table}

\subsection{Determining $I$ and $\Omega$\label{sec:iomega_ic}}

\begin{figure}
\includegraphics[width=2.5in]{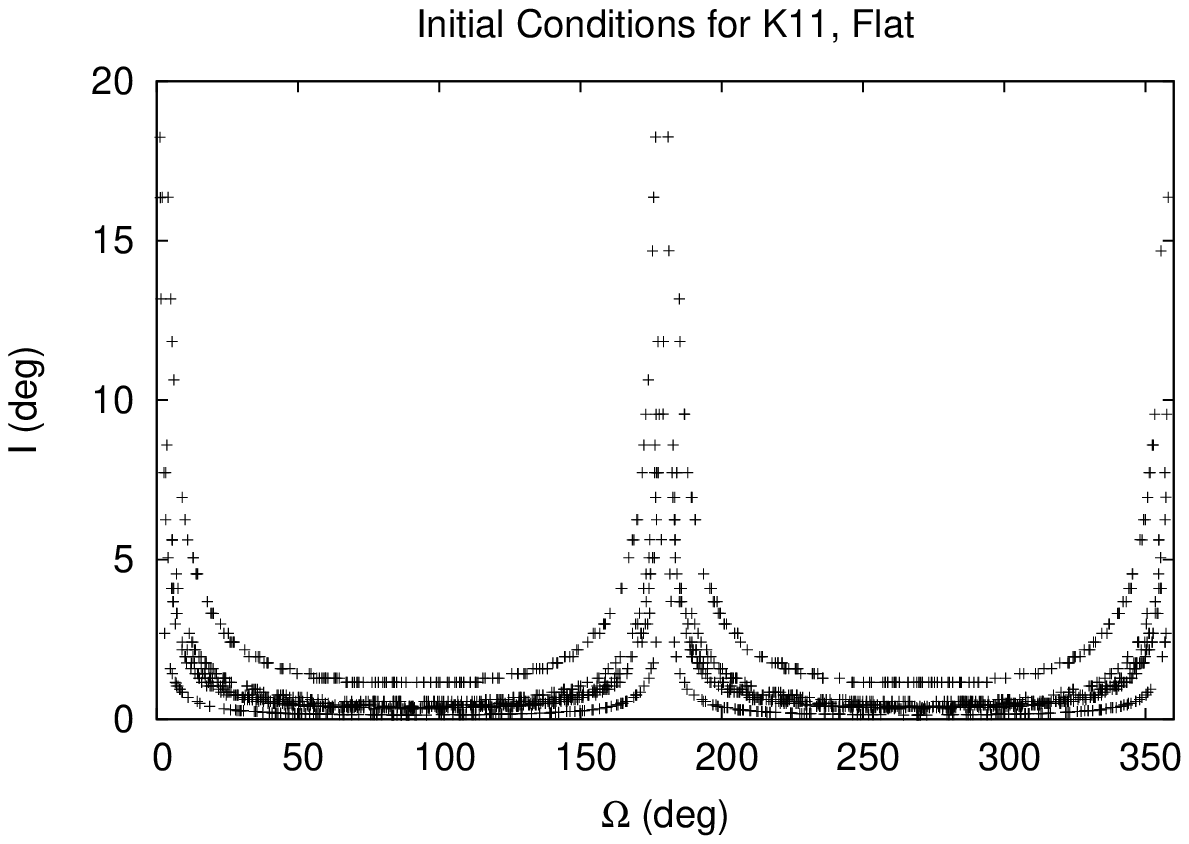}
\includegraphics[width=2.5in]{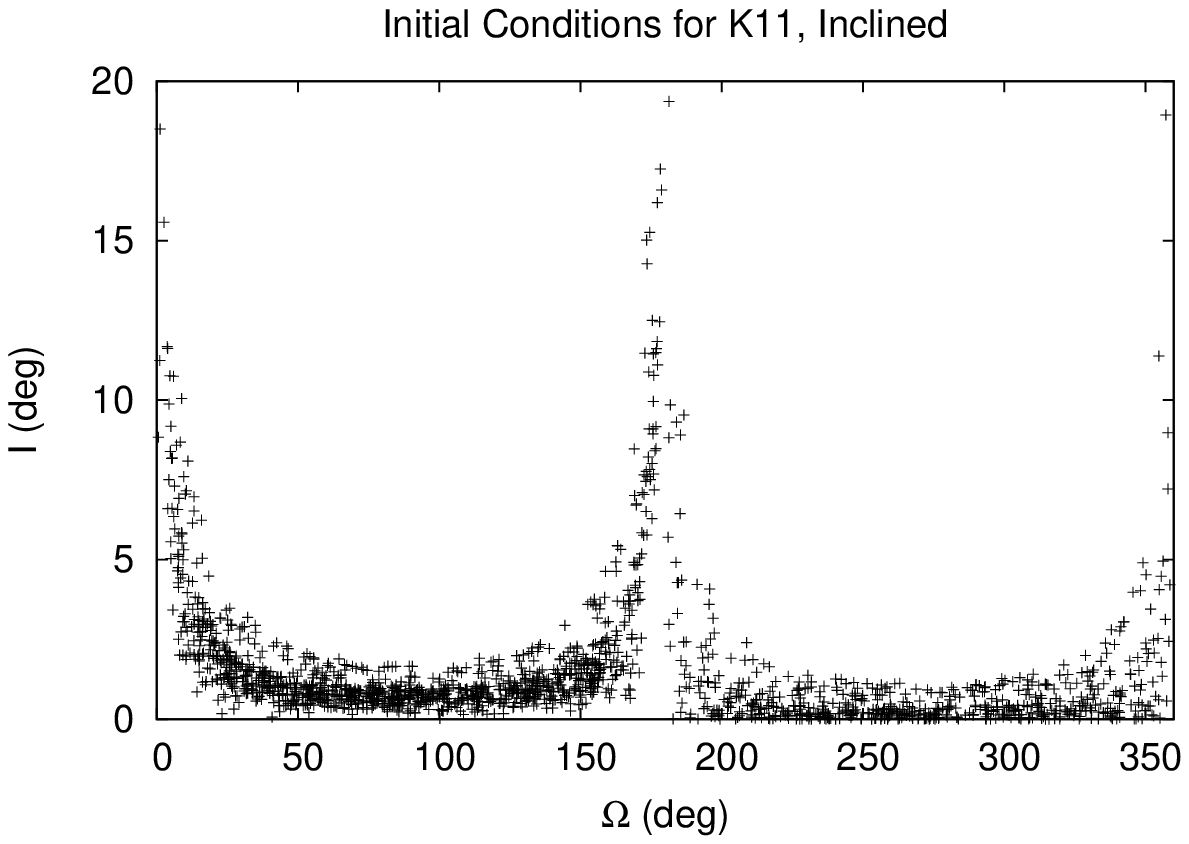}
\includegraphics[width=2.5 in]{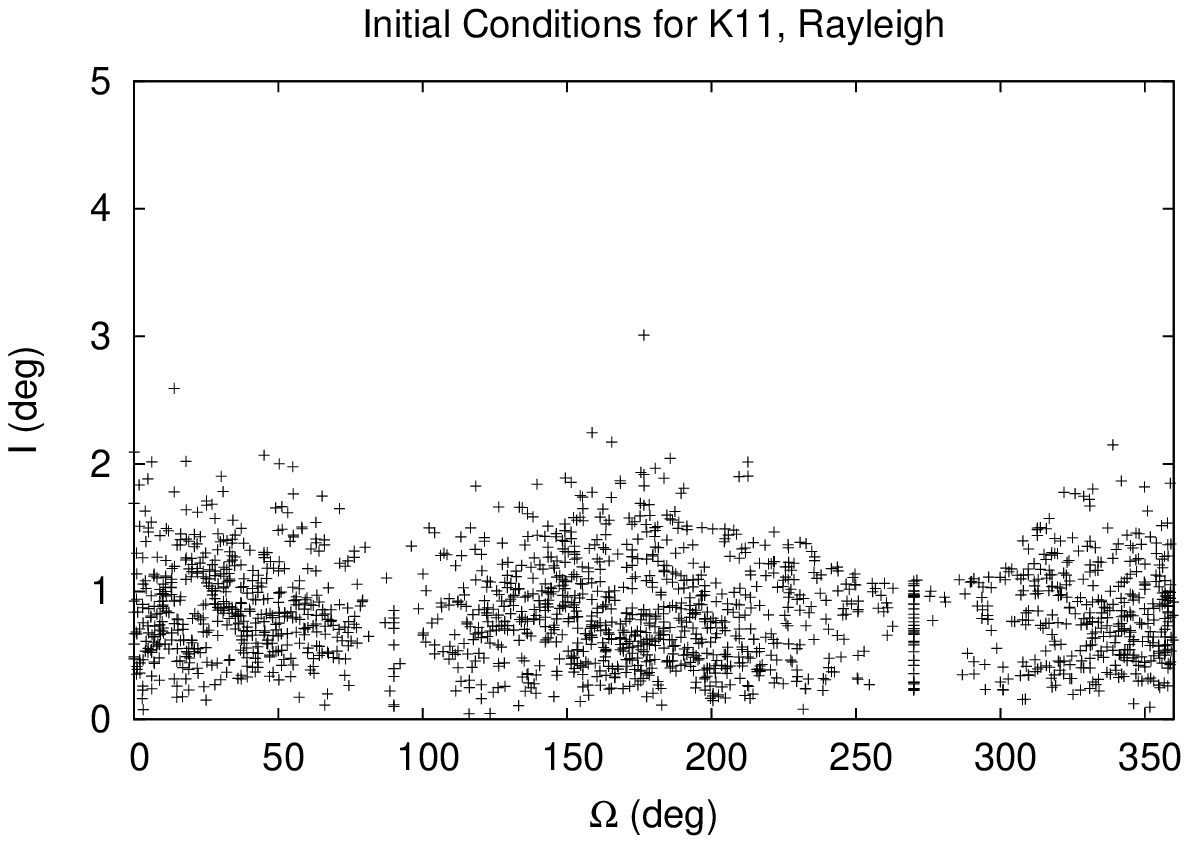}
\caption{Distributions of $I$-$\Omega$ for (left panel) uniform random $\Omega$ with the observer in the orbital reference plane, (middle panel) uniform random $\Omega$ and random observer offset from the orbital reference plane, and (right panel) Rayleigh distributed $i$ with a random observer offset from the orbital reference plane (using a mode of $\sigma = 0.6^\circ$ in this case). } \label{fig:iomega}
\end{figure}

We investigate three scenarios for determining the inclination vectors, which also include different assumptions for a system's orientation relative to the observer.
We refer to these scenarios as flat, inclined, and $\sigma$.  
The first case (flat) assumes that the observer is in the reference plane for the inclinations (this is not necessarily the system's invariable plane).  
As such, an impact parameter $b=0$ means that the orbital inclination of the given planet $I$ is also $0^\circ$, unless the longitude of ascending node is aligned or anti-aligned with the observer's line of sight. 
With this assumption, for a given planet, we draw a random longitude $\Omega$ and then determine an orbital inclination $I$ that would be consistent with the random $\Omega$ and the observationally derived $b$.  
To do this, we assume that the observer lies along the $+x$ axis and that the $y-z$ plane represents the skyplane with the origin centred on the stellar disc.  
The generating script then tries an orbital inclination $I$ for the given $\Omega$ and $\omega$ (where the argument of pericentre $\omega = \varpi-\Omega$), and then determines the minimum $y-z$ distance that a planet on the trial orbit would have from the origin for  $x>0$ (i.e., passing in front of the star).  
If the minimum distance corresponds to a $b$ that is within 5\% of the observationally constrained value (well within the observational uncertainty), then the solution is accepted.  
If not, then a new inclination is drawn as determined from a Newton-Raphson guided search.  
If the script fails to find an inclination within the required tolerance, then the given $\Omega$ is rejected and a new one is tried. 
The nodes are measured counter-clockwise from the $+x$ axis, which means solutions may not exist for $\Omega\approx 0$ or $\pi$ rad when $b>0$.  
An example of an $I$-$\Omega$ distribution for Kepler-11 is shown in Figure \ref{fig:iomega}, left panel.
Each planet (with a corresponding $b$) has a separate solution curve.  

Placing the observer in the reference plane is highly idealized.  
For the second case (the inclined case), we randomly determine an $\Omega$ and then use a trial orbital inclination $I$ to give a proposed orbit.  
The system is then rotated by a random observational inclination $\delta i$ about the $y$ axis, yielding a new skyplane corresponding to the $y-z'$.  
The observer is located along the $+x'$ axis, but is now below the reference plane due to the chosen direction of rotation.  
Here, primes are used to denote the rotated coordinates. 
The procedure as outlined for the flat distribution is then repeated until an impact parameter is found that is within 5\% of the constrained value.  
It should be noted that because the system is no longer at exactly $i=90^\circ$, then $b=0$ could represent an $I>0$, even when $\Omega \ne 0$ or $\pi$ rad. 

The value of $\delta i$ is determined from a uniform random distribution between 0 and $\delta i_{\rm max}$, where $\delta i_{\rm max} = R_{\rm star}/a_{\rm max}$. 
Here, $R_{\rm star}$ is the radius of the star and $a_{\rm max}$ is the semi-major axis of the outermost transiting planet in the given system.  
A system thus has an inclination relative to the observer of $i={\pi}/{2}- \delta i$.
In principle, larger $\delta i$'s than considered here are possible, but this becomes increasingly unlikely with increasing $\delta i$ because it will require very large orbital inclinations among all the planets while maintaining low mutual inclinations. 

A corresponding $I-\Omega$ distribution for Kepler-11 is shown in the middle panel of Figure \ref{fig:iomega} for the inclined distribution.  
The solution curves are similar to the left panel, but there is now much more scatter due to the random observer inclination.  
There is also an offset between inclinations on either side of $\Omega=\pi$. 
Because we have placed the observer below the reference plane and we have chosen $\Omega=0$ to be along $x$ axis, for  $\Omega=[\pi,2\pi)$, an increasing inclination will only increase the impact parameter.  In contrast, for $\Omega=[0,\pi)$, an increasing inclination will first take a given  impact parameter to zero before the impact parameter can increase, thereby allowing a wider range of orbital inclinations.  

Finally, a third case (the $\sigma$ case) is also explored, for which we draw random planet inclinations from a Rayleigh distribution and then determine $\Omega$'s that would be consistent with the given $b$'s.  
The reason for this last approach is to address potential biases in the {\it observed} $\Omega$ distributions. 
A Rayleigh distribution is chosen based on the results of \cite{fabrycky_etal_2014}, in which the mode of the distribution for each system is chosen to be $\sigma=\delta i_{\rm max}$, as defined above.  
The corresponding $\Omega$ is then determined by conducting a gridded search through parameter space, with refinement. 
As in the second set of initial conditions, the observer is randomly positioned to be slightly out of the system's reference plane. 
The resulting $I-\Omega$ distribution, again for Kepler-11, is shown in the right panel of Figure \ref{fig:iomega}.  
Unlike the other distributions, valid system configurations are clustered around $\Omega=0$ and $\pi$.  
There is some striping at $\pi/2$ and $3\pi/2$, which occurs due to certain combinations of $b$, $i$, and $I$.
The striping is likely made a bit more striking due to the $5\%$ solution tolerance and the gridded search.  
They do, nonetheless, represent a narrow region of valid solution space.
The results produced from this third case are critically dependent on $\sigma$.  
To emphasize the effects of a high $\sigma$ on a system, we also run a case for Kepler-11 in which $\sigma = 1.8^\circ$, consistent with the mutual inclination angle derived from the Kepler multiplanet population \cite{fabrycky_etal_2014}. 
For comparison, the $\sigma$ for Kepler-11 as determined by using $\delta i_{\rm max}$ is approximately $0.6^\circ$.

 Before discussing the integrations, we do note that some of the systems, particularly Trappist-1, have strong resonances among planet pairs.  Other than using the semi-major axes and transit timings derived from observations, our  realizations do not enforce that any particular pair of planets is indeed resonant, as determined by the corresponding resonant angle.  As such, some of the realizations may have lower fidelity to the actual systems than others.  Regardless, we are most interested in the variety of transiting behaviours that can occur due to differences in $I$ and $\Omega$ distributions, which can be done with the current simulations.  

\section{Numerical Integration}

We use Rebound with the IAS15 integrator \citep{rein_spiegel_2015} to simulate analogues of Kepler-11, Kepler-445, K2-138, Trappist 1, and Kepler-334 using the initial conditions described in the previous section. 
Each set of initial conditions is run with 300 realizations (i.e., system analogues), and all analogues are integrated for 100 yr.  
The ReboundX extension is used \citep{reboundx} to capture general relativistic effects due to the central body through the ``gr'' option.  

The observer is always assumed to be in the $+x'$ direction, as noted above.  
Each planet's position is recorded at intervals of approximately 5 minutes simulation time ($10^7$ snapshots over 100 yr of simulated evolution).  
Whenever the projected distance $d$ of the planet's centre is within twice the combined radius of the star and the given planet (i.e., $R_{comb} = R_\star + R_p$), the position of the planet is recorded, giving positional information for each transit.  
The beginning of a transit is determined by finding the first recorded point at which the projected distance of the planet from the stellar disc centre $d<R_{comb}$ and linearly interpolating between that time and the point recorded just prior to the planet entering the transit (the ingress for this study).  
The end of the transit is determined the same way using the last point of the series for which $d<R_{comb}$ and the subsequent point out of transit (egress).  
The ingress and egress times are used to determine the duration for the given event.   
The transit midpoint is taken to be halfway through the transit duration. 

\subsection{Stability}\label{sec:stability}

 The above procedures ensure that the initial conditions are reasonably consistent with the actual transit signatures today. However, it does not guarantee that the realizations are stable for the age of the observed systems, which may be of particular importance for the highly inclined cases. We check for stability by running the inclined realizations (the identical initial conditions) for each system up to 100,000 yr and flag each system that has one or more instabilities.  We will ultimately analyze the full set of simulations and then compare that with the outcome of this subset.  The timescale of 100,000 yr was a compromise between completing the simulations in a tractable amount of time while still being useful for understanding the expected long-term instability rate. 

\subsection{Simulation Results}\label{sec:numsim}

The simulations reproduce the initial transit durations for each planet within the uncertainty of the impact parameter.  
The largest deviations are associated with planets that have high initial impact parameters, but overall, the initial transit durations are within about 10\% of the nominal values.  
Using Kepler-11, we verified that the initial conditions also reproduce the nominal initial transit midpoints\footnote{Strictly, provided we start a system realization close to the published reference time, we can recover the expected transit midpoints. For the arbitrary reference JD that we use for all systems, there can be a shift in the transit midpoints relative to published values.}. 
Thus, the system analogues represent plausible configurations despite having a wide range of inclinations for any given planet among the analogues.  

 The overall behaviour of the simulations are summarized in Table \ref{tbl:simsum}. 
In particular, this table shows (1) the percentage of realizations for each system that have at least one de-transiting planet during the 100 year evolution (column 2), and (2) the fraction of analogues that have at least one planet with a TDV $> 10$ min per decade for at least one of the 10 decades (column 4).  

Because some systems have planets on moderate to large initial inclinations (e.g., $>10^\circ$), which can cause a substantial angular momentum deficit, we also analyze subsets of the simulation output, selected by the number of de-transiting planets (zero, one, and two).
By looking at these subsets, we preferentially select system analogues that have different distributions of mutual inclinations (discussed further below).    
With this in mind, column 3 shows the percentage of systems that have one de-transiting planet among all systems that have zero or one de-transiting planet.  Column 5 shows the percentage of analogues that have no de-transiting planets, yet still show a TDV$>10$ min per decade for at least one of the 10 decades. 

It is of interest to understand the extent to which high inclination planets might drive the evolution of a system's transit signatures.  To this end, we find the maximum initial orbital inclination for each system analogue and then determine the median of those maximums.
Hereafter, we call this the median-of-maxes. 
Column six is the median-of-maxes for the zero de-transiting planet subset, column 7 is for one de-transiting planet, and column 8 is for two de-transiting planets.  
The median-of-maxes increases with the number of de-transiting planets, as might be expected.  
Unfortunately, the distributions among the subsets can be large and overlap, providing no clear predictive power.  
We did investigate whether the angular momentum deficit \citep[e.g.,][]{laskar_2008} is a better indicator of a system's potential to have a de-transiting planet.  The results are similar to using the maximum inclinations: each subset has a distribution with a distinct median, but the distributions are broad and overlapping. 
Moreover, the current simulations also do not address whether, given enough time, one of the system analogues with zero de-transiting planets would indeed evolve to a state in which one planet de-transits. 
Future work is needed to determine a metric for evaluating the probability that a given system hosts a de-transiting planet. 

 Before continuing we need to remark on the stability of the systems. As noted in section \ref{sec:stability}, all realizations for each system using the inclined ICs were run for 100,000 yr.  Among K334, K11, K445, K2-138, and Trappist-1, the percentage of realizations that went unstable during that time interval is 0.3\%, 1\%, 5\%, and 9\%, and 20\%, respectively. We expect STIPs to undergo equal fractions of instability in equal decades of time \citep{volk_gladman_2015} after an initial phase of stability.  We very cautiously note that this seems to hold for the Trappist-1 realizations, for which 31 systems decay between $10^3$ and $10^4$ yr of evolution and an additional 29 decay between $10^4$ and $10^5$ yr.  Provided there was nothing peculiar about the realizations that went unstable, should the loss continue at this rate we might expect to have just over half of the realizations left after 1 Gyr.  Too few realizations went unstable in other systems to apply such scaling, and we expect a greater number of realizations to be retained for all other cases.

 For our analysis, if we exclude the systems that go unstable, we see only a small change in the results (a few percent) and only when looking at the full simulation output (shown in Table \ref{tbl:simsum}). The simulations that go unstable are biased toward having many planets de-transit.  For example, among the Trappist-1 realizations that are unstable, all exhibit four or more de-transits in 100 yr.  Because of this, the analyses that consider between zero and 2 de-transits already filter out most of the unstable realizations.

There are two additional points to note from Table \ref{tbl:simsum}.  First, the fraction of systems with de-transiting planets is large, regardless of the initial conditions, except in the cases of K445-flat and all realizations of Kepler-334 (discussed below).
Second, many systems are expected to exhibit variations in transit durations of 10 minutes or more over decade timescales, regardless whether any planets eventually de-transit.  This means that many of the known multiplanet systems should have detectable TDVs.  

For Kepler-445,  there is little possible variation in transit durations if the observer is in the reference plane due to the planets' small impact parameters.   
However, when the system as a whole is inclined relative to the observer (K445-inc or K445-$\sigma$), Kepler-445  shows a behaviour similar to that seen in the other systems.  In stark contrast is Kepler-334.  It shows little variation, with no marked TDVs for any of the realization orientations, including realizations that have highly inclined planets (e.g., $I>10^\circ$).  Based on the secular nodal eigenfrequencies, we should expect this system to have little change in the planets' inclination vectors during the 100 yr integration period, which is borne out by the N-body simulations.

\begin{table}
\caption{Summary of the simulation outcomes. The system names are in the first column, and flat, inclined, $\sigma$ refer to the different initial condition strategies described in section \ref{sec:iomega_ic}. The second column gives the percentage of systems that had at least one planet de-transit during the integration, as compared with all realizations for that system.    The third column is similar, but is the percentage of systems with one de-transiting planet among the systems that had zero or one de-transiting planet during the integration.  The fourth column shows the percentage of systems that had at least one planet with a TDV $>10\rm~min~per~decade$ during any one of the 10 decades. A similar quantity is shown in the fifth column, but only for the systems that  have zero de-transiting planets.  The last three columns give the medians of the maximum inclinations (median-of-maxes) for system realizations that had zero de-transiting planets, that had one de-transiting planet, and had two de-transiting planets, respectively. Kepler-11 (K11) has an additional set of initial conditions, for which planet inclinations are drawn from a Rayleigh distribution with $\sigma = 1.8^\circ$.  Note that all medians for this case are higher than $1.8^\circ$ because the metric only considers the maximum inclination for each system instead of the inclinations of all planets.
 For the inclined initial conditions, some columns report two numbers.  The starred value reflects the results derived by excluding all realizations that go unstable after 100,000 yr. Columns 3 and 5 do not list two values because the simulation subsets with 0 or 1 de-transiting planets contain almost entirely stable realizations, i.e., instability preferentially occurred in systems that had multiple de-transit events.  \label{tbl:simsum}}

\begin{tabular}{llllllll}\hline
System & De-transit  &  De-transit$_{0+1}$ & TDV$>10$ &   $\rm TDV_0>10$&  $\rm Median_0$ & $\rm Median_1$ & $\rm Median_2$\\
 &  \% & \% & \% & \% & $^\circ$ & $^\circ$ & $^\circ$  \\ \hline
K11-flat & 59 &  44 & 93 &  96 & 1.6 & 3.0 & 6.3  \\
K11-inc & 57/57*  & 38 & 98/98* & 95 & 1.9 & 4.3 & 5.7  \\
K11-$\sigma$ & 13  & 12 & 93 & 89 & 1.3 & 1.6 & 1.7  \\
K11-$\sigma_{1.8}$ & 80  & 64 & 100 & 98 & 3.3 & 3.5 & 4.1  \\\hline
K2-138-flat & 83  & 55 & 100 & 98 & 4.7 & 5.7 & 7.7  \\
K2-138-inc & 85/84*  & 59 & 100/100* & 98 & 4.7 & 5.7 & 7.7  \\
K2-138-$\sigma$ & 80  & 66 & 100 & 100 & 4.1 & 4.4 & 5.1  \\\hline
K445-flat & 0.67  & 0.67 & 1.0 & 0.34 & 0.054 & 2.6 & NA  \\ 
K445-inc & 46/43*   & 26 & 62/60*  & 30 & 1.0 & 2.4 & 4.5  \\
K445-$\sigma$ & 63   & 52  & 84 & 55 & 1.9 & 2.4 & 2.9  \\\hline
Trappist-1-flat & 67 & 37 & 78 & 39 & 0.84 & 1.3 & 1.8   \\ 
Trappist-1-inc & 84/80* &47  & 91/89* & 49 &  0.99 & 1.6 & 1.9  \\ 
Trappist-1-$\sigma$ & 69 & 55 & 91 & 72 & 1.0 & 1.1 & 1.2 \\\hline
K334-flat & 0 & 0 & 0 & 0 & 0.93 & - & -   \\ 
K334-inc & 0 &0  & 0 & 0&  3.1& - & -  \\ 
K334-$\sigma$ & 0 & 0 & 0 & 0 & 3.2 & - & -\\\hline

\end{tabular}

\end{table}

\begin{figure}
\includegraphics[width=6in]{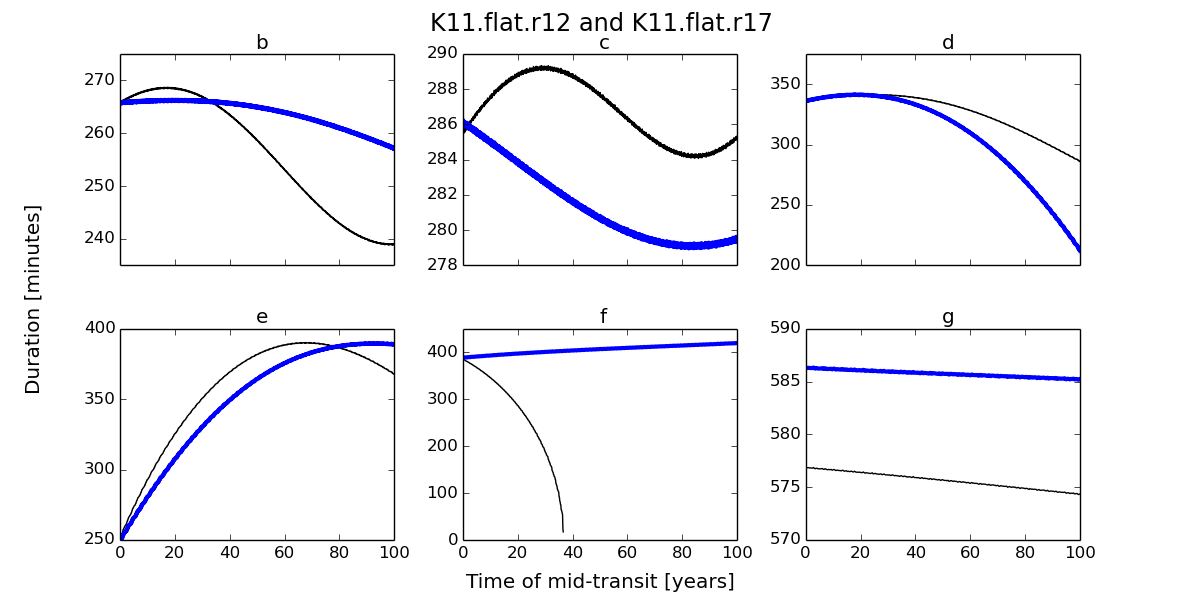}
\caption{Transit durations for Kepler-11 analogues using the ``flat'' initial conditions, with the ``r12'' realization in black and ``r17'' in blue.
There is one planet (planet f) that de-transits during the 100 year evolution in the r12 realization. 
Other realizations can show multiple planets de-transiting within the 100 year time frame. For r12, the maximum initial orbital inclination is $2.4^\circ$.
In contrast, r17 does not have a planet that de-transits, with planets d and e showing the largest overall variation. The maximum initial orbital inclination for this realization is $2.0^\circ$ \label{fig:tdvk11r12}}
\end{figure}

Next, we visualize the behaviour of select system analogues from Kepler-11, Trappist-1, and Kepler-445 to show the complexity of transit duration evolution.  
Figure \ref{fig:tdvk11r12} shows the evolution of two Kepler-11 realizations, one of which has a de-transiting planet (r12, black curves) and the other (r17, blue curve) shows large TDVs, but no de-transiting events.  The names  reflect the realization number used for internal bookkeeping, and in this case, the ICs are from the ``flat'' distribution.
Examining the initial conditions, the maximum inclination among the planets in K11.flat.r12 is $2.4^\circ$, compared with $2.0^\circ$ in K11.flat.r17.  

\begin{figure}
\includegraphics[width=7.5in]{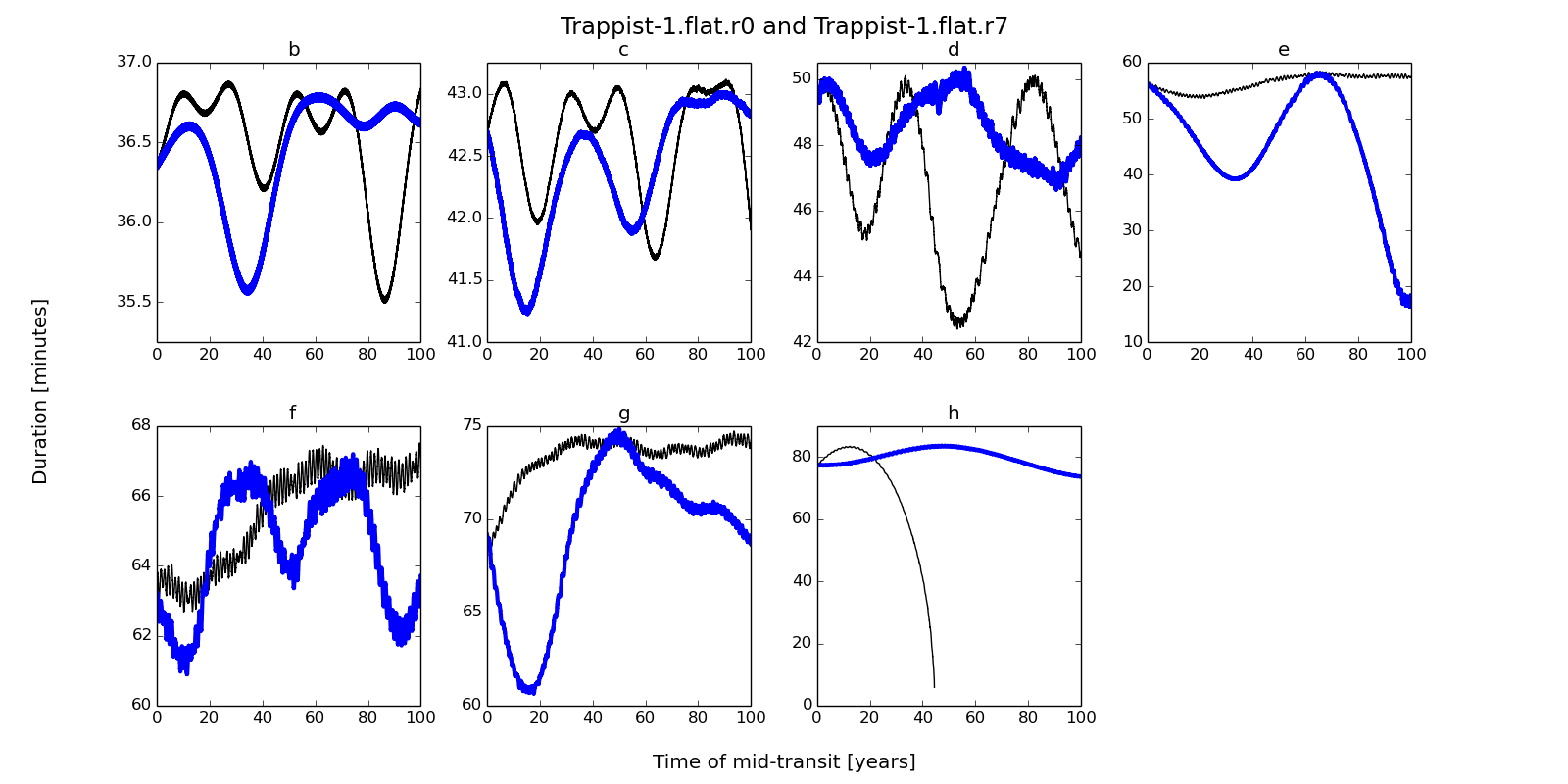}
\caption{Transit durations for Trappist-1 analogues (r0 and r7) using the ``flat'' initial conditions.
Planet h de-transits after 40 years of evolution for r0 (black curve). In r7 (blue curve), no planets de-transit, although planet e comes very close to doing so toward the end of the simulation. \label{fig:tdvtrapr0}}
\end{figure}

A similar behaviour is seen in Trappist-1.  
The analogue Trappist-1.flat.r0 (Fig.~\ref{fig:tdvtrapr0}) has one de-transiting planet, while analogue Trappist-1.flat.r7 (also Fig.~\ref{fig:tdvtrapr0}) has only large TDVs for the duration of the simulation.  
Again, the actual transit duration variation signatures are diverse, and we are highlighting two cases as examples. Looking at the initial conditions of these realizations, the maximum inclination among the planets in r0 is $1^\circ$, while it is $0.8^\circ$ for r7.  
Although the inclination difference alone is small, the three-dimensional orientation of the planets will also be different between realizations, which will play a role in the observational outcome.

\begin{figure}
\includegraphics[width=6in]{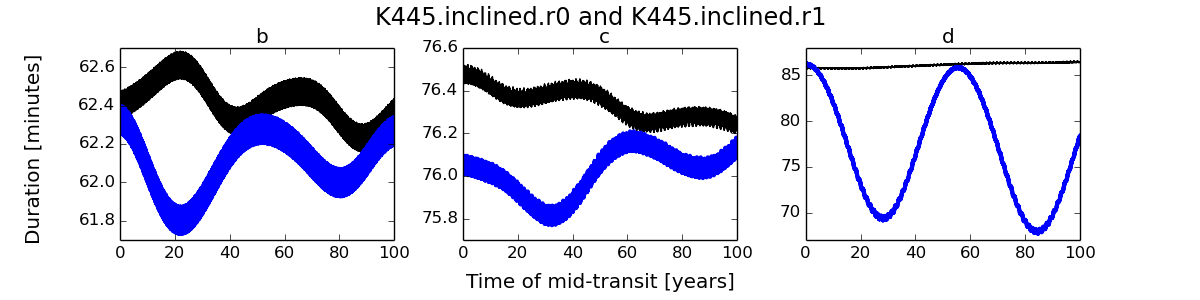}
\caption{Transit durations for Kepler-445 analogues using the ``inclined' initial conditions.
The TDVs are all small for r0 (black), although they do show longterm trends. In contrast, one planet shows large TDVs on decade timescales for realization r1 (blue).  The maximum inclination for r1 is about $1.2^\circ$, while that for r0 is $0.3^\circ$. \label{fig:tdvk445r0}}
\end{figure}

As a final example, we show in Figure \ref{fig:tdvk445r0} two realizations from K445 using ICs from the ``inclined'' distribution.  Both cases exhibit no de-transiting planets, but have very different TDV signatures.  The system K445.inclined.r0 has a maximum inclination of $0.3^\circ$, while the system in Figure K445.inclined.r1 has a maximum planet inclination  of about $1.2^\circ$.  

Altogether, the transit duration evolution of planets in any given system has many possible trajectories, but these can be constrained over time. 

\section{Discussion}\label{sec:discussion}

As many of the Kepler STIPs are approaching a decade since their discovery, measuring TDVs for a large number of systems is a feasible observational goal, provided the TDVs are of the magnitude found here. 
As noted in the introduction, a number of TDVs have already been measured, and we expect many more to be announced in the upcoming years. 

We reiterate that the impact parameter inferred from observations does not directly correspond to a planet's inclination.  
As seen from possible initial conditions in Figure \ref{fig:iomega}, different positions of the ascending node permits distributions of orbital inclinations to remain consistent with a given set of observed transit durations.  
Taking all the realizations for all ICs (i.e., flat, inclined, and $\sigma$) of the N-body simulations at face value, we find that de-transiting planets should be common in STIPs with high nodal eigenfrequencies.  Kepler-11, Kepler-445, K2-138, and Trappist-1 all have at least one set of initial conditions in which over 50\% of the realizations have a planet that de-transits. 
These same systems, excluding the special case of K445-flat,  have over 10\% of the analogues produce a de-transiting planet for all IC distributions.  

The ICs that showed the fewest occurrences of de-transiting planets include K11-$\sigma$, K445-flat, and all of the Kepler-334 ICs.  The K11-$\sigma$ IC distribution used $\sigma\approx 0.6^\circ$ in the Rayleigh distribution for the inclinations (see section \ref{sec:iomega_ic}), and as such, had few planets with even moderately high inclinations (see the median-of-maxes in Table \ref{tbl:simsum}).  
The planets in the K445-flat set of ICs  showed little transit duration evolution among almost all realizations.  
Although the nodal eigenfrequencies are fast for this system, the magnitudes of the inclination vectors must be small, in this case, due to the combined effect of having small impact parameters ($b\approx 0$) and having the observer in the reference plane.    
Allowing a random position for the observer (K445-inc) opened the potential 3D orbital configuration space, and de-transiting planets were again possible, although at a reduced rate compared with other systems.  

Kepler-334 showed little variation in transit durations for all realizations.  This particular STIP has some of the slowest eigenfrequencies of all the systems we explored, and as discussed above, we did not expect it to show strong TDVs.  The N-body simulations of Kepler-334 highlight the utility of secular theory precession rates for identifying good targets for TDV follow up. 

As a check for consistency between the actual precession rates and expectations from secular theory, we reran the inclined realizations for all systems in Table \ref{tbl:icsims}, but did so for only 1 yr and tracked the orbital element evolution.  This was used to get instantaneous nodal precession rates for each planet in each realization.  Because the actual precession rate may vary throughout a precession period, these instantaneous rates do show a distribution of values (very approximately Rayleigh distributed) with a clear mode.  Looking at just the magnitudes of the maximum precession rates for each realization, the modes  are approximately (in degrees per year)
9.0, 8.0, 3.5, 1.2, and 0.1 for Kepler-445, Trappist-1, K2-138, Kepler-11, and Kepler-334, respectively.
The actual precession rates for a planet are a linear combination of the different eigenfrequencies, so we do not expect an exact correspondence between the two.  Nevertheless, the magnitude of the maximum precession rates are within approximately 30\% of the maximum eigenfrequencies, suggesting we can indeed use nodal eigenfrequencies as an indicator for potential TDVs.  

A major caveat for interpreting these results is that we assumed that all configurations consistent with the observations are equally possible, including very large mutual inclinations between the planets.  
This again is the motivation for subsetting the simulation output in Table \ref{tbl:simsum}.  Should independent, physical reasons  (e.g., near planar configurations in a gaseous disc with limited subsequent excitation) dictate  that actual orbital inclinations in STIPs remain low (such as below the values in column 6 of Table \ref{tbl:simsum}), then de-transits may be much less abundant than that found here. 
However, column 7 of Table \ref{tbl:simsum} shows that the systems with one de-transiting planet do not have maximum inclinations that are obviously  too high.  Likewise, our understanding of high-multiplicity systems might be biased not only by seeing systems with low inclinations, but possibly by STIPs with planets on moderate inclinations and chance alignments of nodes. 

Even among the subsets for which there were no de-transiting events, TDVs $>10$ minutes on decade timescales is very common and shared among all the different initial condition realizations.  
Kepler-11, K2-138, and TRAPPIST-1 are all good candidates for exhibiting strong TDVs.  
For general prioritization of long-term monitoring of STIPs, we suggest using nodal precession frequencies, as determined from secular theory.  Systems that also have planets with large impact parameters are particularly good targets (e.g., Table \ref{tbl:rank_first}).

If we take STIPs with eigenfrequencies of $-1^\circ~\rm yr^{-1}$ or faster as a rough cut for TDV systems of interest (Kepler-11 is just faster than this threshold), then of the 118 systems in Tables \ref{tbl:rank_last} through \ref{tbl:rank_first}, 41 are candidates. 

This work is different but complementary to that by \cite{becker_adams_2016}, who also used secular theory but focused on changes to the magnitude of the orbital inclination rather than the full inclination vector.  
{\it This creates a potentially large difference in our results regarding the expectation for systems that might de-transit} (at least for those that we studied), where we find that de-transiting planets (or newly appearing transits) should be common over decade timescales. 
 For example, Becker \& Adams found that systems remain in a state such that their planets are transiting $>85$\% of the time. While we focus on the percentage of systems that have a de-transiting event rather than the time, our results cannot be immediately reconciled.  Instead, the higher incidence of de-transiting events could be due to our consideration of the full inclination vector rather than just its magnitude, making the Becker \& Adams results a lower limit on the de-transit rate.   

 A more detailed comparison between the studies can be done by comparing the number of recorded transits, summed over all planets and realizations, with the corresponding expected number of transits over the 100 yr evolution.   This calculation thus gives us a sense of the average fraction of time that a planet will remain transiting  for all realizations.  Using only the simulations that are stable over 100,000 yr (only the inclined ICs), the corresponding fractions are 100\%, 91\%, 86\%, 85\%, and 80\% for K334, K11, K445, Trappist-1, and K2-138, respectively.  However, these percentages are misleading, because they are averages over all planets.  If, instead, we ask what is the fraction of time that all realizations have all planets transiting based on the planet that spends the most time out of transit (if one exits), then the fractions are 100\%, 65\%, 74\%, 52\%, and 53\% again for  K334, K11, K445, Trappist-1, and K2-138, respectively. The actual fraction of time that at least one planet will be out of transit could be lower still, as we have not taken into account having multiple planets go out of transit at different times.  These results are comparable to those reported in Table \ref{tbl:simsum}, showing that we do indeed find a higher rate of de-transits than Becker \& Adams. 
Nonetheless, while the de-transiting rates are different between the studies, the general magnitudes of the TDV calculations agree that TDVs $>10$ min per decade should be common.

Finally, we note that STIPs among, say, the Kepler sample are snapshots of evolving planetary systems  and do not include additional planets, should they exist at larger orbital periods.  Systems with both transiting and non-transiting planets \citep[such as Kepler-20;][]{buchhave_etal_2016} may represent cases in which secular cycling has caused a relatively recent de-transit event or cases in which a new transiting event should be expected to appear,  as has occurred for K2-146 \citep{hamann_etal_2019}.  Efforts to refine the transit properties of planets \citep[e.g.,][]{goldberg_etal_2019,christ_etal_2019} will further help to reveal the presence of TDVs. 

\section{Summary} \label{sec:summary}

We have investigated whether transit duration variations are observable on decade timescales by using secular theory (118 STIPs) and direct N-body integration (5 STIPs).  
The N-body simulations  took into account possible variations of the full inclination vectors when creating system analogues.  
Three different methods were used for creating initial conditions, with each method containing 300 realizations that are observationally consistent with the corresponding STIP.      

For the STIPs we investigated using N-body simulations, tens of percent or more of the realizations contained at least one planet that de-transited, i.e., the secular cycling of their inclination vectors took the planet out of a transiting state. 
At least some of the short-period systems that are known should exhibit de-transiting (or even a new planet coming into transit) through long-term monitoring.

The fastest nodal eigenfrequency, as determined from secular theory, is a predictor for which systems could have strong TDVs. 
Systems that have eigenfrequencies of $-1^\circ ~\rm yr^{-1}$ should be prioritized (41 of the 118 systems we examined meet this threshold), particularly if their planets also have large impact parameters.  Eight of the 118 systems we examined have eigenfrequency faster than $-3^\circ~\rm yr^{-1}$ and may be of particular interest.    
Measuring the resulting TDVs of these systems could reveal their planets' 3D planetary orbital orientations and/or constrain the presence of additional planets in the system.

The authors thank D.~Ragozzine  and the anonymous referee for comments that improved this manuscript. This work was supported in part by an NSERC Discovery Grant, The University of British Columbia, The Canadian Foundation for Innovation, and the BC Knowledge Development Fund. This research was undertaken, in part, thanks to funding from the Canada Research Chairs program.

\bibliography{bib}

\end{document}